\documentclass{articlestyle}

\usepackage{lineno,hyperref}
\modulolinenumbers[5]

\begin{document}

\begin{frontmatter}

\title{Distributed data processing and analysis environment for neutron scattering experiments at CSNS}

\author[mymainaddress,mysecondaryaddress]{H.L. Tian}

\author[mymainaddress,mysecondaryaddress]{J.R. Zhang\corref{mycorrespondingauthor}}
\cortext[mycorrespondingauthor]{Corresponding author}
\ead{jrzhang@ihep.ac.cn}

\author[mymainaddress,mysecondaryaddress]{L.L. Yan}

\author[mymainaddress,mysecondaryaddress]{M. Tang}

\author[mymainaddress,mysecondaryaddress]{L. Hu}

\author[mymainaddress,mysecondaryaddress]{D.X. Zhao}

\author[mymainaddress,mysecondaryaddress]{Y.X. Qiu}

\author[mymainaddress,mysecondaryaddress]{H.Y. Zhang}

\author[mymainaddress,mysecondaryaddress]{J. Zhuang}

\author[mymainaddress,mysecondaryaddress]{R. Du}

\address[mymainaddress]
{China Spallation Neutron Source(CSNS), Institute of High Energy Physics (IHEP), Chinese Academy of Sciences (CAS), Dongguan 523803, People¡¯s Republic of China}
\address[mysecondaryaddress]
{Dongguan Institute of Neutron Science(DINS), Dongguan 523808, People¡¯s Republic of China}

\begin{abstract}
China Spallation Neutron Source (CSNS) is the first high-performance pulsed neutron source in China, which will meet the increasing fundamental research and technique applications demands domestically and overseas. A new distributed data processing and analysis environment has been developed, which has generic functionalities for neutron scattering experiments. The environment consists of three parts, an object-oriented data processing framework adopting a data centered architecture, a communication and data caching system based on the C/S paradigm, and data analysis and visualization software providing the 2D/3D experimental data display. This environment will be widely applied in CSNS for live data processing.
\end{abstract}


\end{frontmatter}

%\linenumbers

\section{Introduction}

With the rapidly increasing volume of high quality experimental data provided by recent neutron scattering instruments, computing infrastructure that supports data processing and analysis is significantly pressurized. Meanwhile, the instrument software needs to hide the complex nature of experimental data manipulation, including data acquisition, reconstruction, reduction, analysis and visualization. At China Spallation Neutron Source (CSNS) \cite{Wang:2013aka}, raw data collected by a Data Acquisition System (DAQ) is, first, reconstructed to produce neutron events that constitute  hit position and time-of-flight (TOF). Then, those neutron events with related meta-data are converted to a NeXus \cite{Konnecke:po5029} file that stores experimental data based on both event \cite{Peterson201524,Zhao2011107} and histogram modes, and other experimental information. Consequently, reduction software analyzes four-dimensional structure (X, Y, Z and TOF) of measured data to momentum and energy transfer representing the properties of the samples. The analysis and visualization software provides graphical user interface, and enables the users to have a better understanding of the collected data during the experiment.  Other environments for this include ADARA \cite{adara} at ORNL,  Manyo-Lib \cite{Suzuki2004175} at J-PARC,  DAVE \cite{DAVE} at NIST, GumTree \cite{Rayner20061333, Lam20061330} at OPAL, etc.  The packages of software for data processing and visualization include ISAW \cite{Tao2006422} at IPNS, Mantid \cite{Arnold2014156} at ISIS/SNS, LAMP \cite{LAMP} at ILL, etc.

There is a variety of neutron scattering instruments, representing different operating principles, designs and implementations. Therefore, the architecture of the instrument software must be reusable and flexible to benefit the efficiency of software development and maintainability of the data processing infrastructure. Our aim is to develop a generic data processing framework that provides basic computing functionalities, a common communication system that defines interaction between distributed components, and graphical user interface modules using a model-view-controller (MVC) design pattern to fit the instruments at CSNS. Additionally, a distributed computing environment should be available on heterogeneous platforms, including Windows, Linux and Mac OS X.

The data processing framework written in C++ results in high-performance handling of large-scale data in an object-oriented approach. The framework also provides Python APIs by BOOST.Python \cite{boost}, which enables users to develop their own application based on this framework without any time-consuming building procedure.

The communication system and graphical user interface are designed in Python, which offers rapid development using rich Python libraries and the capability to invoke the Python APIs of the data processing framework. The communication system provides asynchronous messaging, a naming and routing service, a data caching service and remote procedure call (RPC). The data caching service is employed in the graphical user interface as a model that is responsible for managing the client data in the MVC design pattern. This approach reduces the data-related server load. RPC and messaging are used to interact with other distributed components in the data processing and analysis environment. Naming and routing services provide the location transparent access, which is convenient for the distributed components to communicate. The distributed data processing and analysis environment is developed to satisfy the requirements of the neutron scattering experiments at CSNS and other facilities. Here we describe each of the modules to give an overview of the environment.

\section{Data processing framework - DroNE}

DroNE (Data pROcessing suit for Neutron Experiments) was originally designed to support online and off-line data processing for the neutron scattering experiments at CSNS. The development of DroNE is based on SNiPER \cite{Zou:2015ioy}, which is a light weight framework to support multi-type physical experiments at the Institute of High Energy Physics (IHEP), such as the Jiangmen Underground Neutrino Observatory (JUNO) \cite{Djurcic:2015vqa}, Large High Altitude Air Shower Observatory (LHAASO) \cite{DiSciascio:2016rgi} and CSNS.

The design of DroNE is inspired by Gaudi \cite{BARRAND200145} and Mantid \cite{Arnold2014156}. The framework provides a number of generic components and standard mechanisms for a general experimental data processing task. Two major categories of components encapsulated in DroNE, are algorithms and services. An algorithm will manipulate certain types of input data objects, and produce new data objects as output. A service offers  essential functionalities that are directly or indirectly required by algorithms. The framework maintains directories of algorithms and services by managers, AlgMgr and SvcMgr respectively, which allow other components to locate the requested algorithm or service by a name.

\begin{figure}
  \centering
  \includegraphics[width=6cm]{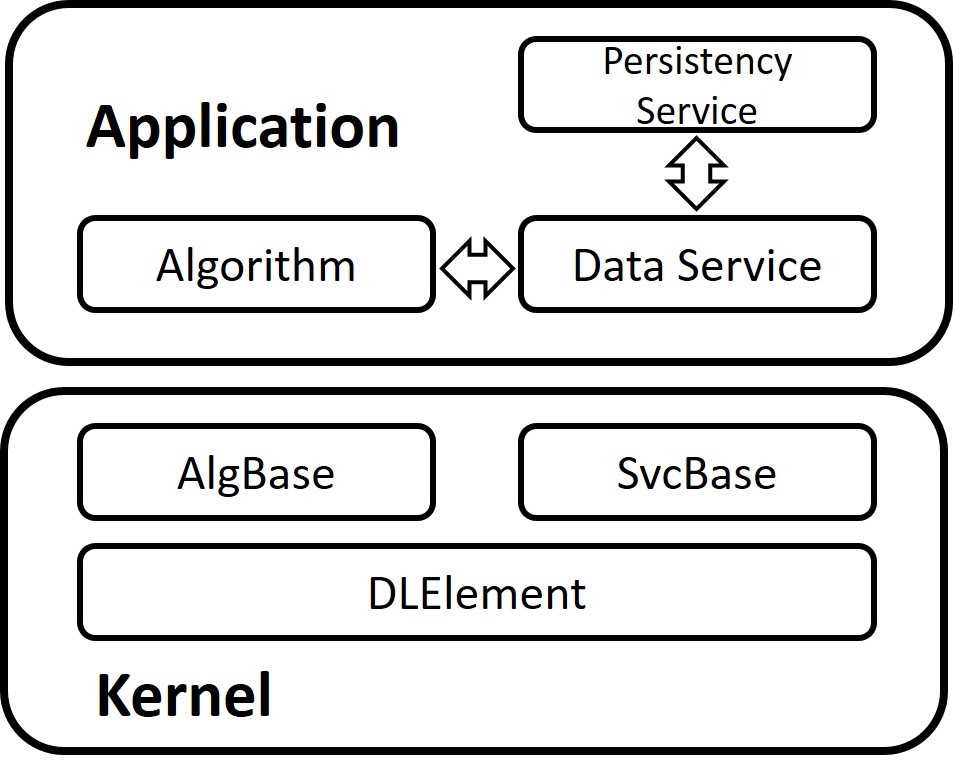}\\
  \caption{Diagram of the DroNE framework. The DroNE kernel can be accessible by users to support different data processing applications.}\label{DRONE}
\end{figure}

The architecture of DroNE is shown in Figure.\ref{DRONE}. The components in the DroNE kernel are functional elements with a well-defined public interface, have specified relationships, and default interactions with other components. Most of the components are inherited from a common base class DLElement (Dynamic Loadable Element), in which a finite-state machine is used to manage and track the run-time status.  Two of the most important components in the kernel are AlgBase and SvcBase, which are virtual base classes defining abstract methods of the algorithms and services. A class template named DLEFactory implements the factory method design pattern \cite{DesignPatterns}.  Both AlgFactory and SvcFactory  are aliases of an instantiation of the DLEFactory, which are responsible for creating the concrete instances of the algorithms and services according to the class name. This design supports dynamic component discovery and generic component factories, that allow users to implement plug-and-play data processing applications.

DroNE has been designed to be easily customized when used by different data processing applications. The customization will be in terms of new data processing algorithms and new data structures. As shown in Figure.\ref{DRONE}, those "user codes" are encapsulated inside two main locations: algorithms and Data Service. Algorithms are developed by scientist programmers, who have adequate knowledge in a specific scientific domain and experience in programming. To provide an easy method for the algorithms to access data in a transparent manner, DroNE adopts a data centered architectural style, which also enables the development of algorithms in an independent way. Data Service implemented by experienced framework and application developers, provides a hierarchical transient data store, which offers simplified data access to the algorithms. The Data Service has a "tree" structure with one "root" node, in which "leaf" nodes and "branch" nodes are registered. Leaf nodes, called "DataObject", contain concrete data, such as values, matrixes, events, in addition to others. The others are "branch" nodes, which  have one or more sub-branch/DataObject. This design is very useful to store a large number of data objects into one Data Service. Using Data Service as a means of communication between algorithms, data objects can be identified by their logical path, and the coupling between independent algorithms is minimized.

Algorithms do not directly access the data objects in the data file. Two approaches are adopted to populate the transient data stores from persistent data and vice versa. One is called the  Persistency Service that sustainably converts a specific data object from its persistent representation into its transient one or the other way around. The other is that, instead of implicit data loading and saving processing, explicit data handling algorithms are involved to load and save data.

\begin{figure}
  \centering
  \includegraphics[width=8cm]{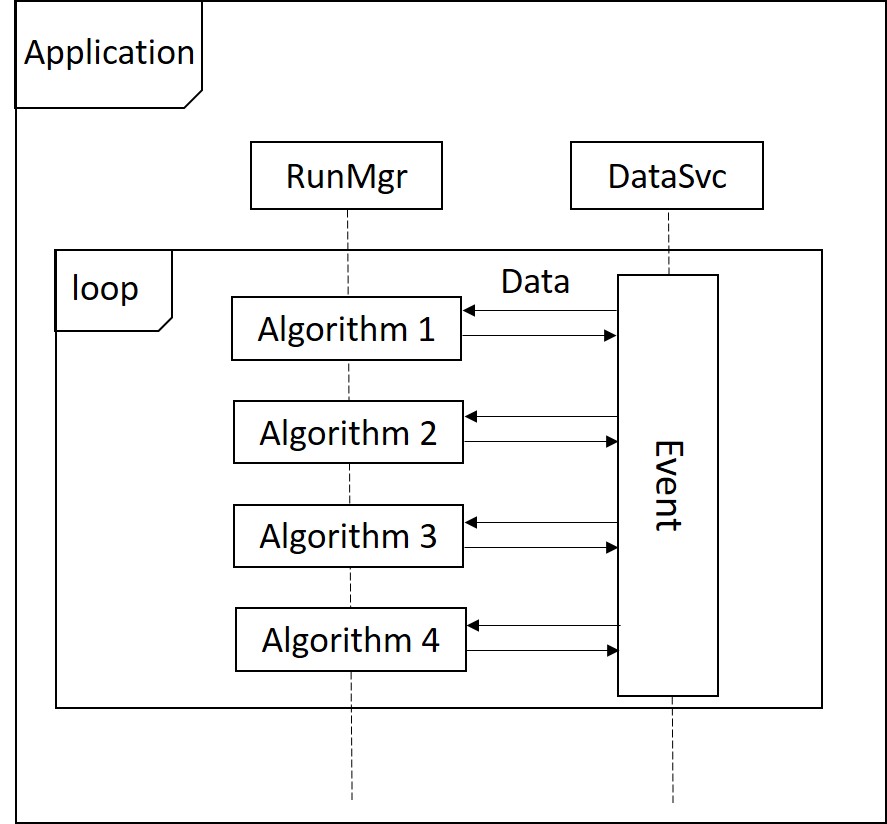}\\
  \caption{Event loop sequence diagram of DroNE.}\label{EVTLOOP}
\end{figure}

\begin{figure}
  \centering
  \includegraphics[width=8cm]{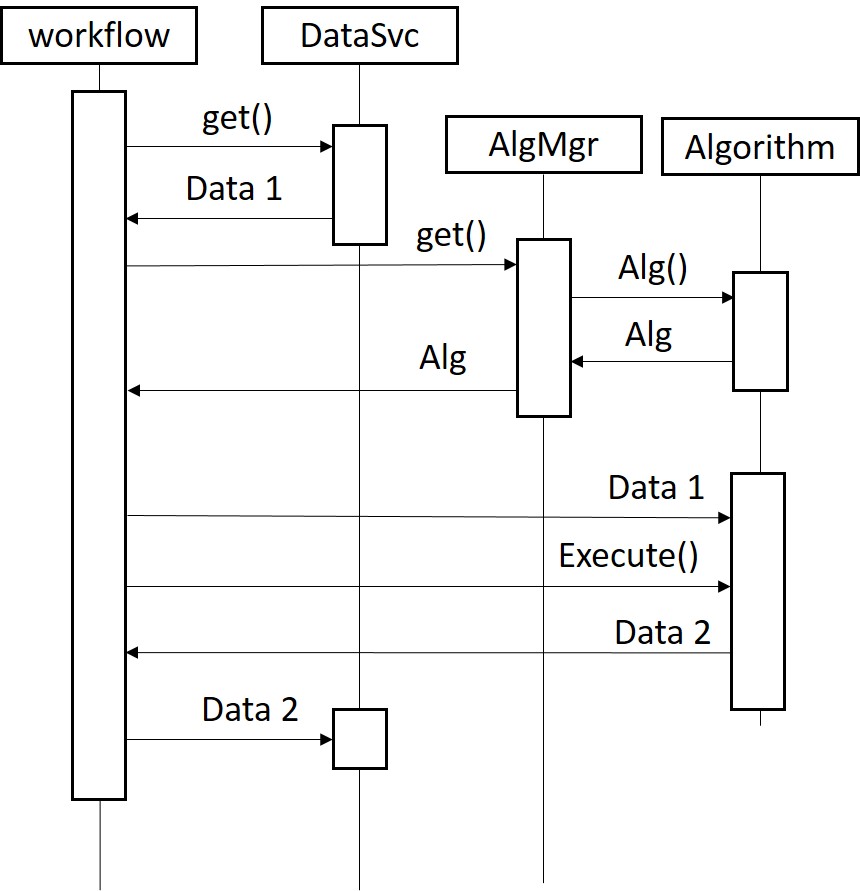}\\
  \caption{Work flow sequence diagram of DroNE.}\label{WORKFLOW}
\end{figure}

Event loop and work flow are both implemented in this framework. In the event-loop mode as shown in Figure.\ref{EVTLOOP}, a RunMgr of the application initializes all the algorithms and services. Then, a DataSvc sets up the collection of events that will constitute the basis of its iteration (loop) and the various transient storages by using the Persistency Service. Moreover, a regular algorithms scheme is invoked to process each event.
In the work flow mode shown in Figure.\ref{WORKFLOW}, the data processing chains are employed by a work flow engine, which is developed in Python to perform an algorithm dynamic invocation according to the context. Data objects from the Data Service will be taken as input data for an algorithm, and a new data object will be stored within the Data Service that contains the results of the algorithm.

\section{Communication and data caching system - NEON}

NEON (Neutron Experiment Over Network) is designed to provide a communication layer to all distributed components involved by CSNS data processing tasks,  especially the live data processing and analysis system. The live data processing and analysis system is responsible for raw data reconstruction, experimental data reduction, analysis and monitoring, graphical user interface, etc. Basic functionalities are fulfilled to improve the efficiency and reliability of the application, including data caching, naming and routing services for location transparency, and asynchronous communications.

\begin{figure}
  \centering
  \includegraphics[width=8cm]{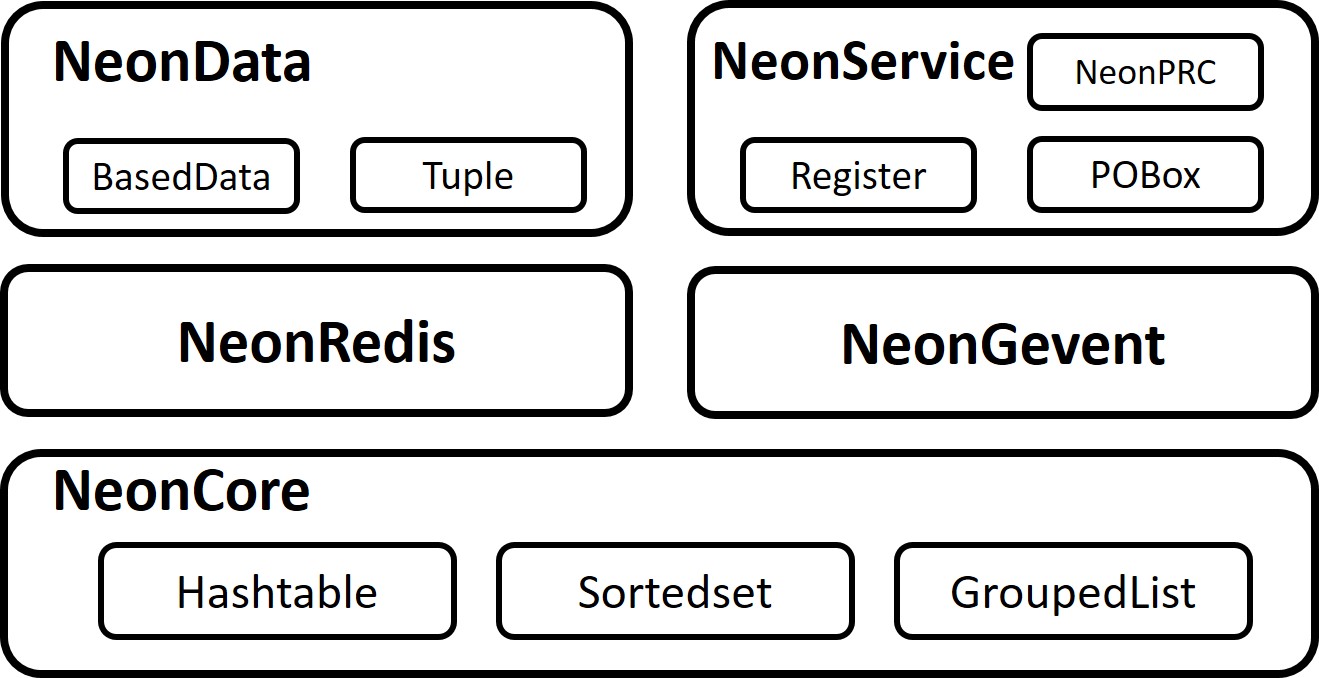}\\
  \caption{Design of NEON. The NeonService is a collection of services exchanging commands and information. The NeonData provides data caching functionalities.}\label{NEON}
\end{figure}

Two basic APIs in NEON are NeonData and NeonService as shown in Figure.\ref{NEON}. NeonData provides a simple and efficient data caching service. The server and client can synchronize the contents through a central data server. Two methods are employed in each NeonData object. One is "dump" which uploads the local data to the central data server, and the other is "load" which  downloads the data from the central data server into the local cache. Two basic types of NeonData are implemented: "BasedData" and "Tuple". The BasedData does not store historical data in the central data server, in which the data is refreshed when 'dump' is executed. However, Tuple keeps historical data in the central data server to provide log information. The two types of data can be derived to support user defined data structures.

NeonService is another key feature, which provides "Register", "POBox" and "NeonRPC" services. The Register service is introduced in order to allow for the required transparency. NEON instances publish their names and access addresses by registering them with the Register service; then, clients can query the access address by the name of the instance. Two access approaches are employed in NEON. The first is that the instance runs as a server, and waits for requests from the clients. In this direct mode, the access address contains the IP and port of the server. The second is that the central data server is used as a message broker, from which instances regularly fetch messages. In this indirect mode, the access address is simply a logical path pointing to the data storage within the central data server. The POBox service provides a messaging service. The instance publishes its access address in the Register service, and can receive messages sent from other distributed components in direct or indirect modes. The differences between those two modes are transparent to the high-level application.

A distributed application often applies Remote Procedure Calls (RPC) to build a collaborative environment for coordinate  operations among different components. JSON-RPC \cite{jsonrpc} is one of the lightweight remote procedure call protocols, which uses JSON as the data format to communicate information between distributed components. According to the JSON-RPC specifications, a remote method is invoked by sending a request to a remote service, and unless the request is a notification, it must receive a response. However, this synchronous mechanism is heavy and ill-suited for DroNE. For most of the DroNE applications, the data processing task executes in a mono-threaded manner. As long as data streams are received or APIs are invoked, DroNE automatically pipelines the separate requests, no matter whether the requests use blocking or asynchronous access. Therefore, DroNE does not offer a synchronous response mechanism under this consideration. Another reason is that NEON does not favor bi-directional communication in order to decouple the location of the client and server.

\begin{figure}
  \centering
  \includegraphics[width=10cm]{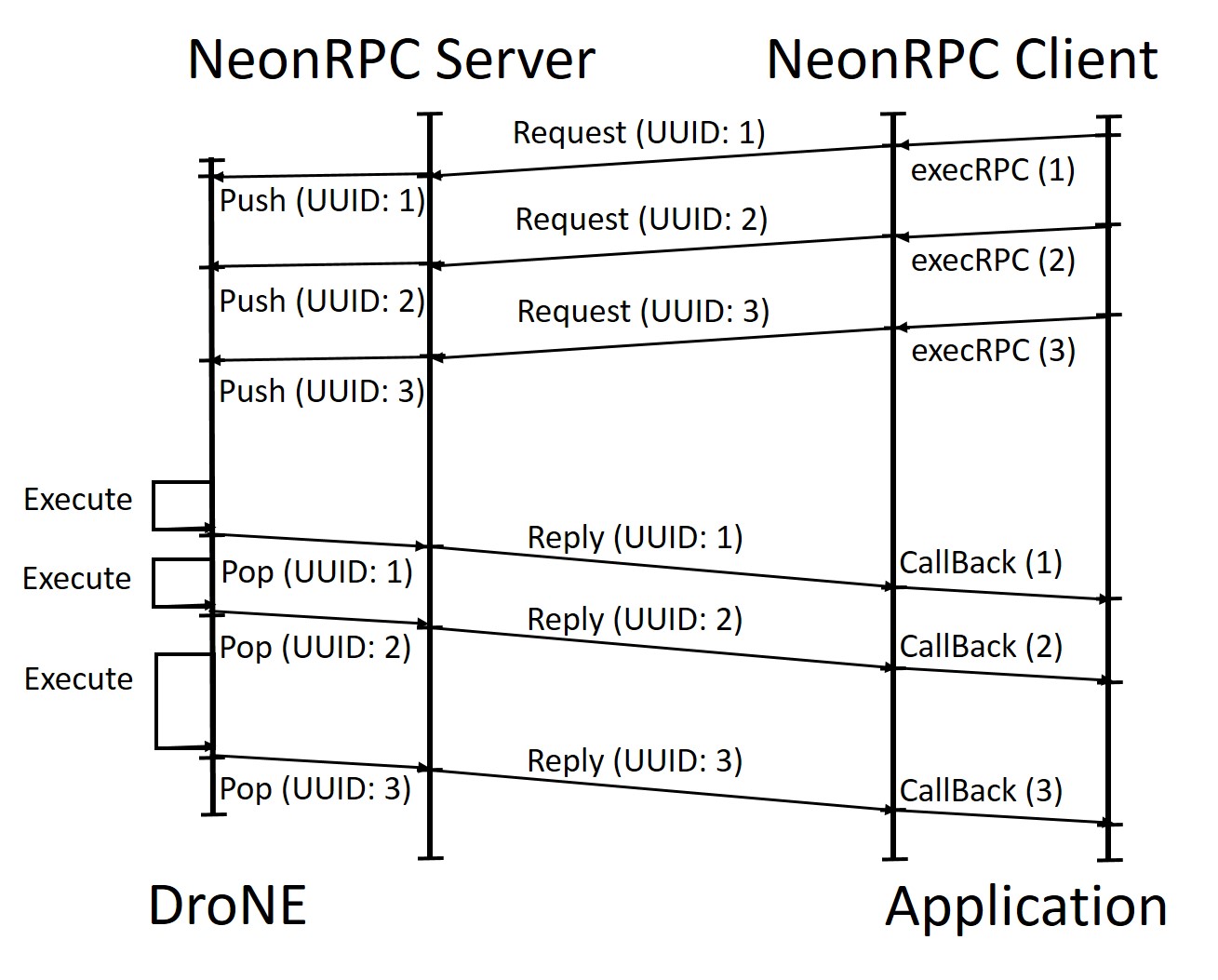}\\
  \caption{NeonRPC asynchronous mechanism. The client does not have to wait for the reply and proceeds with its task.}\label{RPC}
\end{figure}

NeonRPC is based on JSON, but includes slight differences in the implementation. Without the requirement of an instant response, the asynchronous NeonRPC mechanism as shown in Figure.\ref{RPC} is implemented over the NEON POBox service. The NeonPRC client keeps a UUID (Universally Unique Identifier) when sending a RPC request to the remote service, and registers a callback function associated with the UUID. The request is a single Python dictionary serialized by JSON, which contains "method", "parameters", and "id". The "id" has three properties: "to"-the name of the remote service, "from"-the name of the client, and "uuid"-the UUID identifying the invocation. The POBox sends this RPC message according to the "to" property. When an RPC method is received, the server also keeps this UUID and pushes the invocation into the pipeline of execution. After the method call is completed, the server replies with a response according the UUID. The response is a Python dictionary serialized by JSON as well, which contains "result", "error", and "id". The "id" of the response is created automatically according to the UUID kept by the server, and also possesses three properties: "to"-the name of the client, "from"-the name of the server, and "uuid" , same as that in the RPC request. Finally, the response is received by the client. Then, the client invokes the callback function according to the UUID. Moreover, NEON is a pure Python application without considering multi-language support; therefore, the "parameters" of the request and the "result" of the response support the Python-specific data format by Pickle \cite{pickle}, which is a Python library provides the capability to  serialize and de-serialize a Python object structure.

\begin{figure}
  \centering
  \includegraphics[width=10cm]{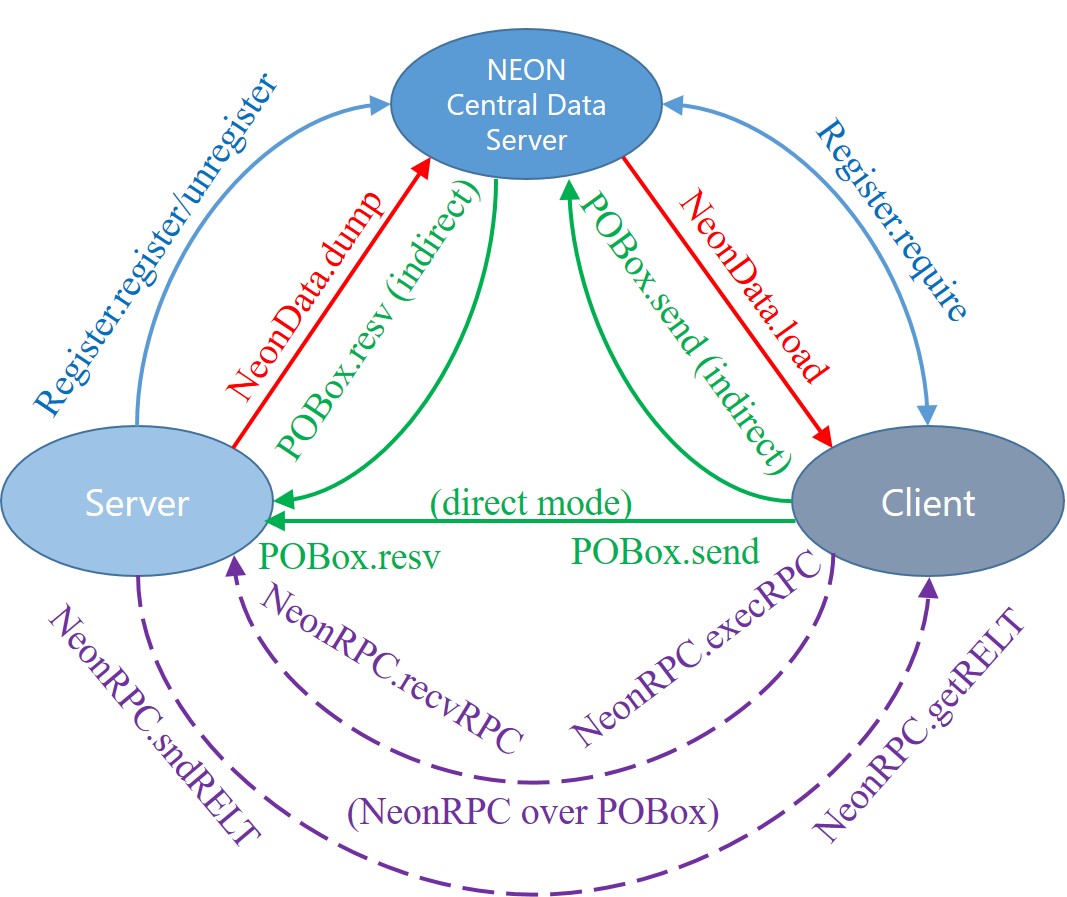}\\
  \caption{Main data flow diagram of NEON.}\label{DFD}
\end{figure}

Using the C/S mechanism and a central data server, the interaction between the different components of NEON is shown in Figure.\ref{DFD}. A server is a process that publishes information in the central data server or receives messages from clients. A client is any process that consumes the information produced by a server, or sends messages to a server.  All information about servers are handled by the Register service, and can be requested by the client. The server also has the capability to perform the RPC method based on the POBox messaging service, which can be called by any client. The methods of NEON and their implementations shown in Figure.\ref{DFD} are detailed in the discussion above.

The implementation of the central data server and integration technology that provides the glue between the distributed components is not specified in NEON.  NeonRedis is one of the central data server implementations for NEON as shown in Figure.\ref{NEON}, which uses the in-memory data store Redis \cite{redis} as a database. NeonGevent provides point-to-point communication APIs on top of Gevent \cite{gevent} - a coroutine-based Python networking library. Other SQL/NoSQL database and integration technologies, like MySQL and RabbitMQ,  will be considered.

Users feel free to choose their own technique, provided the following specified data structures and methods are supported, such as Hashedtable, Sortedset and Groupedlist pre-defined in NeonCore, as shown in Figure.\ref{NEON}. Groupedlist is the base of the POBox service, where two methods, "pull" and "push", are required. For Hashedtable and SortedSet, which are the bases of the Register service and NeonData, the "request" and "publish" methods are required. The importance of standardizing the interfaces guarantees a smooth integration and reuse of high level applications.

\section{Data analysis and visualization framework - DataPilot}

\begin{figure}
  \centering
  \includegraphics[width=12cm]{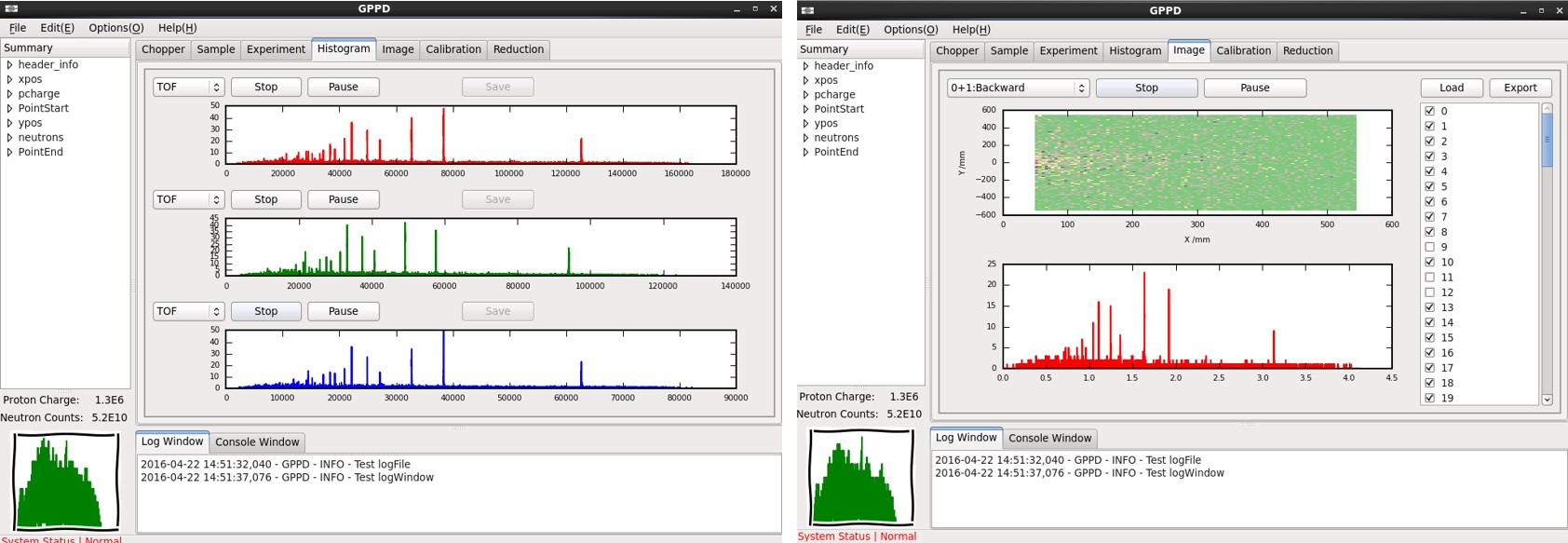}\\
  \caption{Screen shots of DataPilot showing the visualization of the histograms and detector counts.}\label{PILOT}
\end{figure}

DataPilot, the data analysis and visualization framework developed in Python, is designed to display the real-time data streaming. Figure.\ref{PILOT} shows the DataPilot graphical interface, which provides visualization functionality for instrument, sample and experiment status, data histograms and detector images on a tabbed panel, as well as abilities to calibrate, reduce and analyze the experimental data. DataPilot communicates with DroNE and the EPICS (Experimental Physics Industrial Control System) \cite{epics} control server using NEON and pyEpics \cite{pyepics}, which enables access measured data in DroNE and Process Variables (PV) in EPICS. The histograms and curves are generated automatically to assess the status of the experiment and determine the operation followed in the experiment. Many types of graphs are introduced to monitor the experimental status and sample environment, including 2D detector images, histograms (Intensity vs lambda, Intensity vs Q, etc.) and curves. Algorithms covering peak-finder, curve fitting and other arithmetic operations, can perform scientific analysis of experimental data.

\section{An application: Live data processing and analysis environment at CSNS}

\begin{figure}
  \centering
  \includegraphics[width=12cm]{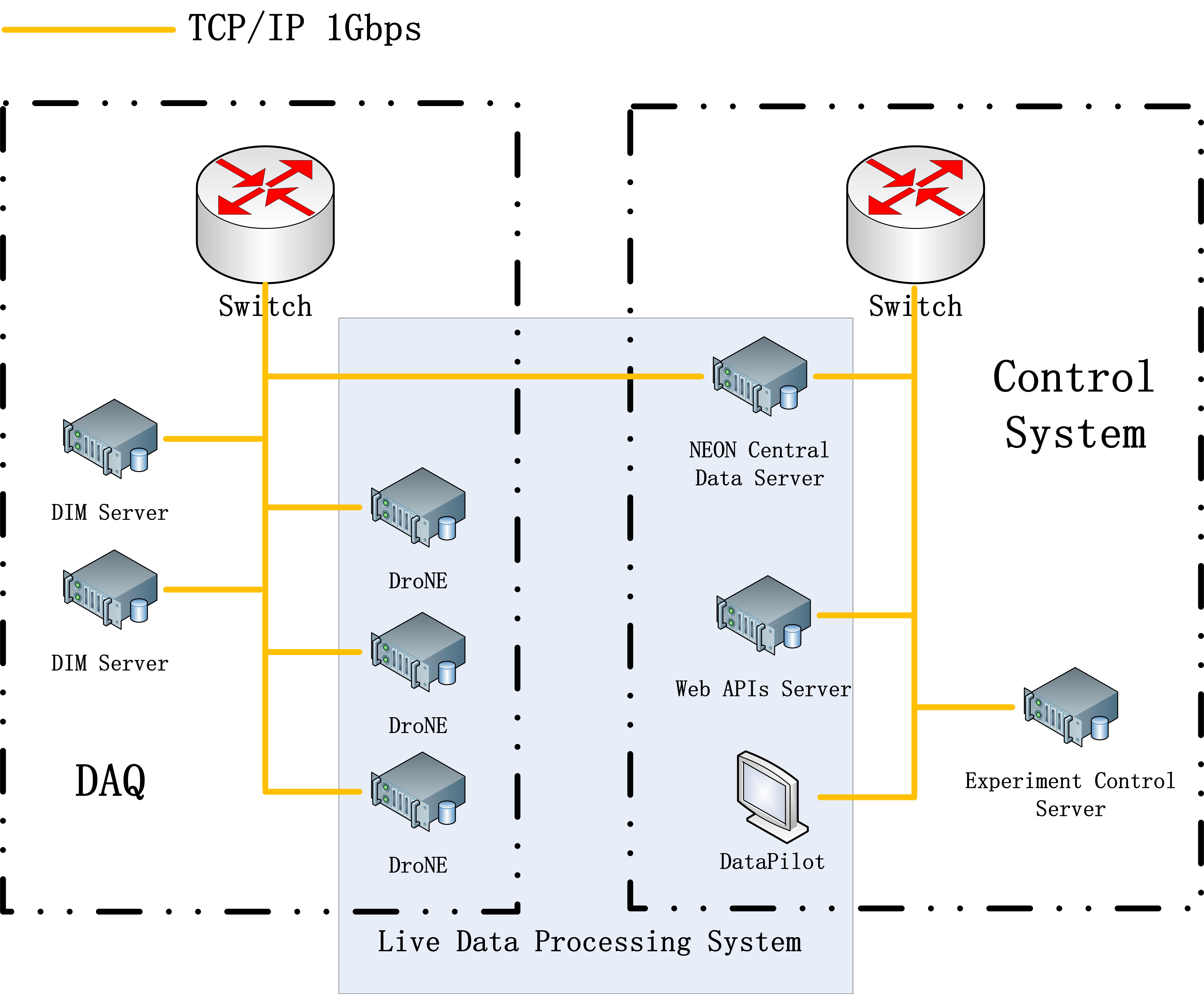}\\
  \caption{Topology of the live data processing and analysis system at CSNS}\label{TOPOLOGY}
\end{figure}

\begin{figure}
  \centering
  \includegraphics[width=12cm]{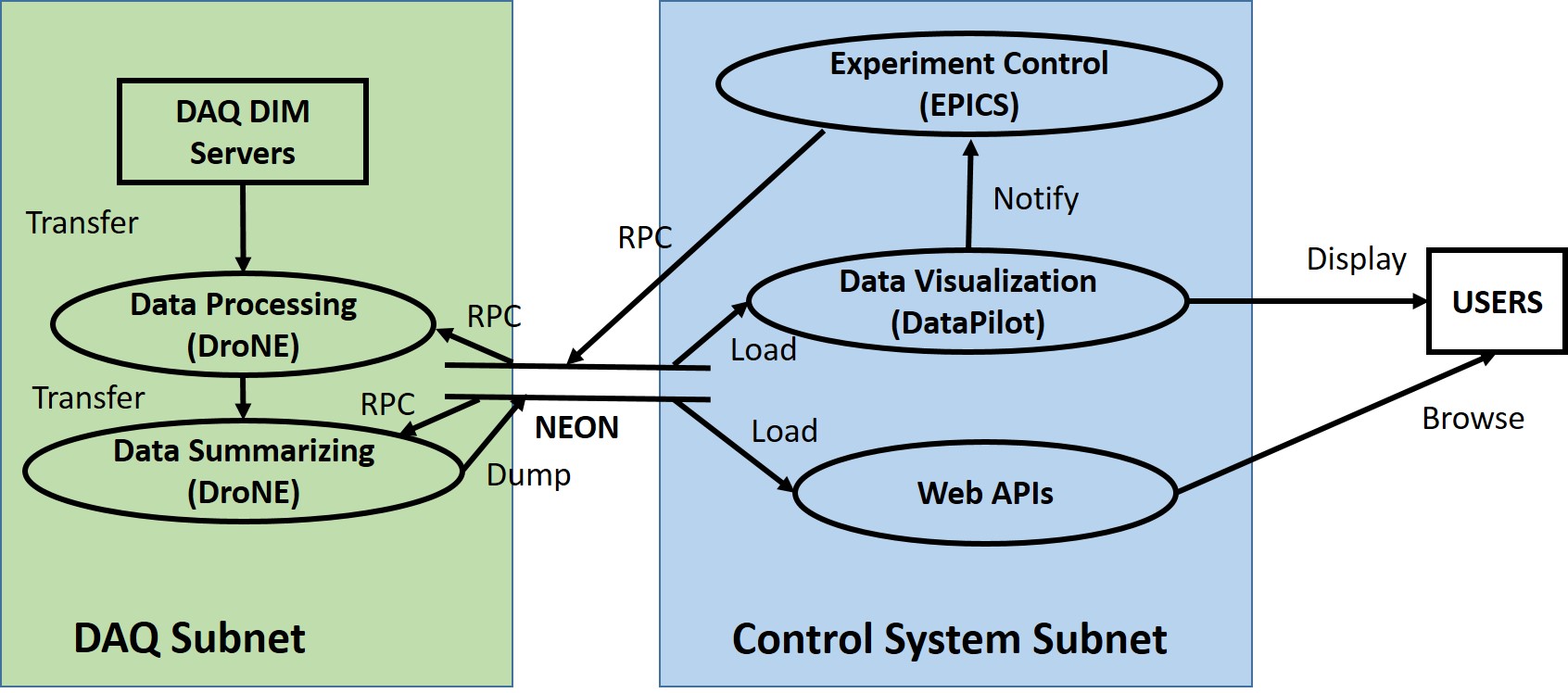}\\
  \caption{Data-flow diagram of the data processing procedure for CSNS. }\label{GPPD}
\end{figure}

The topology of the environment applied to the live data processing and analysis system for neutron instruments at CSNS is shown in Figure.\ref{TOPOLOGY}. There are two parts in this environment, one is the data processing part running in the DAQ subnet, the other is the data visualization part running in the control system subnet. The NEON central data server is an information agent, which has dual-network cards to respond to the connections from the DAQ subnet and the control system subnet. Figure.\ref{GPPD} shows the schematic analysis flow diagram for live data processing. The event data collected by the DAQ is transferred to an online data processing system by DIM \cite{dim}, when measured data are reconstructed. Then, reconstruction data are merged into histogram objects by data summarizing software. Those two components are both based on DroNE. NEON DataObjects stored in the central data server are employed as intermediaries that transmit experimental data  across subnet boundaries. DataPilot is introduced to offer users a simple manner to analyze  those experimental data. Web APIs provide a person with the capability to access the status of instrument operation from outside. PVs are sent to EPICS to notify the control system of the change in experimental status. The control system invokes the RPC methods to start/stop the procedure of data processing and analysis.

\section{Conclusion}

 The distributed data processing and analysis environment for neutron scattering experiments has been developed. DroNE is implemented with an object-oriented methodology in C++, which offers an easy method to add and update functions with high performance. Combined with Python, DroNE makes it easier for the scientists to contribute advanced codes. NEON provides the environment with the capacity of scalability and interactivity. DataPilot offers  a simple manner to access and analyze the experimental data. This environment providing common data processing, analysis, visualization, caching and transmission functionalities, is applied to the data processing tasks in CSNS.

\section*{Acknowledgements}
This work is supported by National Natural Science Foundation of China (No. 11305191).
We would also like to thanks all users for their suggestions, especially the DAQ Group and Control System Group at CSNS.

\bibliography{mybibfile}

\end{document}